\documentclass[conference]{IEEEtran}
\IEEEoverridecommandlockouts
\usepackage{cite}
\usepackage{amsmath,amssymb,amsfonts}
\usepackage{algorithmic}
\usepackage{graphicx}
\usepackage{textcomp}
\usepackage{xcolor}
\usepackage{multirow}
\def\BibTeX{{\rm B\kern-.05em{\sc i\kern-.025em b}\kern-.08em
    T\kern-.1667em\lower.7ex\hbox{E}\kern-.125emX}}
\begin{document}

\title{SAR ATR under Limited Training Data Via MobileNetV3
\thanks{This work was supported by the National Natural Science Foundation of China under Grants 61901091 and 61901090.}
}

\author{
\IEEEauthorblockN{Chenwei Wang, Siyi Luo, Lin Liu, Yin Zhang, Jifang Pei, Yulin Huang, and Jianyu Yang}
\IEEEauthorblockA{\textit{School of Information and Communication Engineering} \\
\textit{University of Electronic Science and Technology of China}\\
Chengdu, China}
}

\maketitle

\begin{abstract}
In recent years, deep learning has been widely used to solve the bottleneck problem of synthetic aperture radar (SAR) automatic target recognition (ATR). 
However, most current methods rely heavily on a large number of training samples and have many parameters which lead to failure under limited training samples. In practical applications, the SAR ATR method needs not only superior performance under limited training data but also real-time performance.
Therefore, we try to use a lightweight network for SAR ATR under limited training samples, which has fewer parameters, less computational effort, and shorter inference time than normal networks. At the same time, the lightweight network combines the advantages of existing lightweight networks and uses a combination of MnasNet and NetAdapt algorithms to find the optimal neural network architecture for a given problem. 
Through experiments and comparisons under the moving and stationary target acquisition and recognition (MSTAR) dataset, the lightweight network is validated to have excellent recognition performance for SAR ATR on limited training samples and be very computationally small, reflecting the great potential of this network structure for practical applications.
\end{abstract}


\section{Introduction}
Synthetic aperture radar (SAR) has been of major importance in civil and military applications because of its ability to provide high-resolution, weather-independent images throughout the day \cite{intro1,wang2019parking,wang2021deep}. 
Automatic Target Recognition (ATR) is able to use computer processing power to predict the class of targets. For this reason, SAR ATR has been of great interest and is very challenging \cite{intro2}.
Recently, many excellent scholars have proposed various deep learning-based methods with good results in SAR ATR applications \cite{intro3,intro4,intro5,wang2023entropy,wang2023sar,wang2022semi,wang2020deep,wang2022sar,wang2021multiview,wang2020multi}.

However, these methods always require each target type to contain sufficient labeled training samples to achieve good performance, which is often difficult to achieve in the real world. 
In practice, the SAR measurement process has a high cost and targets of interest may be located in inaccessible areas. What’s more, in earthquake rescue and maritime rescue, the number of available SAR images may be only less. This means that the quantity of the images would be limited to train the model and thus lead to performance degradation. 

Aiming to solve the SAR ATR problem under limited training data, a lot of research has been presented \cite{intro6,intro7,intro8,EliMRec,SLMRec,BundleGT,DBLP:conf/acl/Zhao00WZZC23,DBLP:conf/emnlp/ZhaoWLSZ022,wang2022recognition,wang2022global}. 
For example, Wang \textit{et al} \cite{intro9} proposed a novel few-shot learning framework named hybrid inference network (HIN) for SAR ATR based on a CNN for embedding feature projection. Besides, in order to extract azimuth-insensitive features in a SAR ATR task with only a few training samples, they \cite{intro10} designed a convolutional bidirectional long short-term memory (Conv-BiLSTM) network as an embedding network to map the SAR images into a new feature space where the recognition problem became easier. Furthermore, Cao \textit{et al} \cite{intro11} developed a label-directed generative adversarial networks (LDGAN) based on the Wasserstein distance, and generated more realistic SAR target images to help improve the recognition performance in the presence of data shortage. 

Although these aforementioned studies show good potential with extremely limited training images, in real-world applications, real-time performance is also a major concern in addition to recognition effect considerations. Therefore, considering the impact of complex network structure on computation time, we try to find a lightweight network for SAR ATR.

From 2016 until now, the industry has proposed lightweight network models such as SqueezeNet \cite{SqueezeNet}, ShuffleNet \cite{ShuffleNet}, NasNet \cite{NasNet}, MnasNet \cite{MnasNet}, and MobileNet \cite{MobileNet}, which make it possible to run neural network models on mobile terminals and embedded devices. And MobileNet is more representative of lightweight neural networks, especially MobileNetV3  \cite{MobileNetV3} launched by Google. On the ImageNet dataset, its variant MobileNetV3-Large achieves a Top-1 accuracy of 75.2\%.

In this paper, we apply the representative lightweight network MobileNetV3 to SAR ATR and confirm its outstanding recognition efficiency with limited data samples through experiments on the moving and stationary target acquisition and recognition (MSTAR) dataset. Meanwhile, to verify the real-time performance of the lightweight network, we quantitatively applied the floating points operations (FLOPs) measure to compare it with other SAR ATR methods, and the results show that the lightweight network has excellent real-time performance while guaranteeing an outstanding recognition rate.

The rest of this paper is organized as follows. Section \uppercase\expandafter{\romannumeral2} describes the framework of the lightweight network MobileNetV3-Large and the baseline architecture of some modules. Section \uppercase\expandafter{\romannumeral3} shows the experimental results of the lightweight network for SAR ATR and comparison with other methods. Finally, Section \uppercase\expandafter{\romannumeral4} draws a short conclusion. 

\begin{figure*}
\centering
\includegraphics[width=0.88\textwidth]{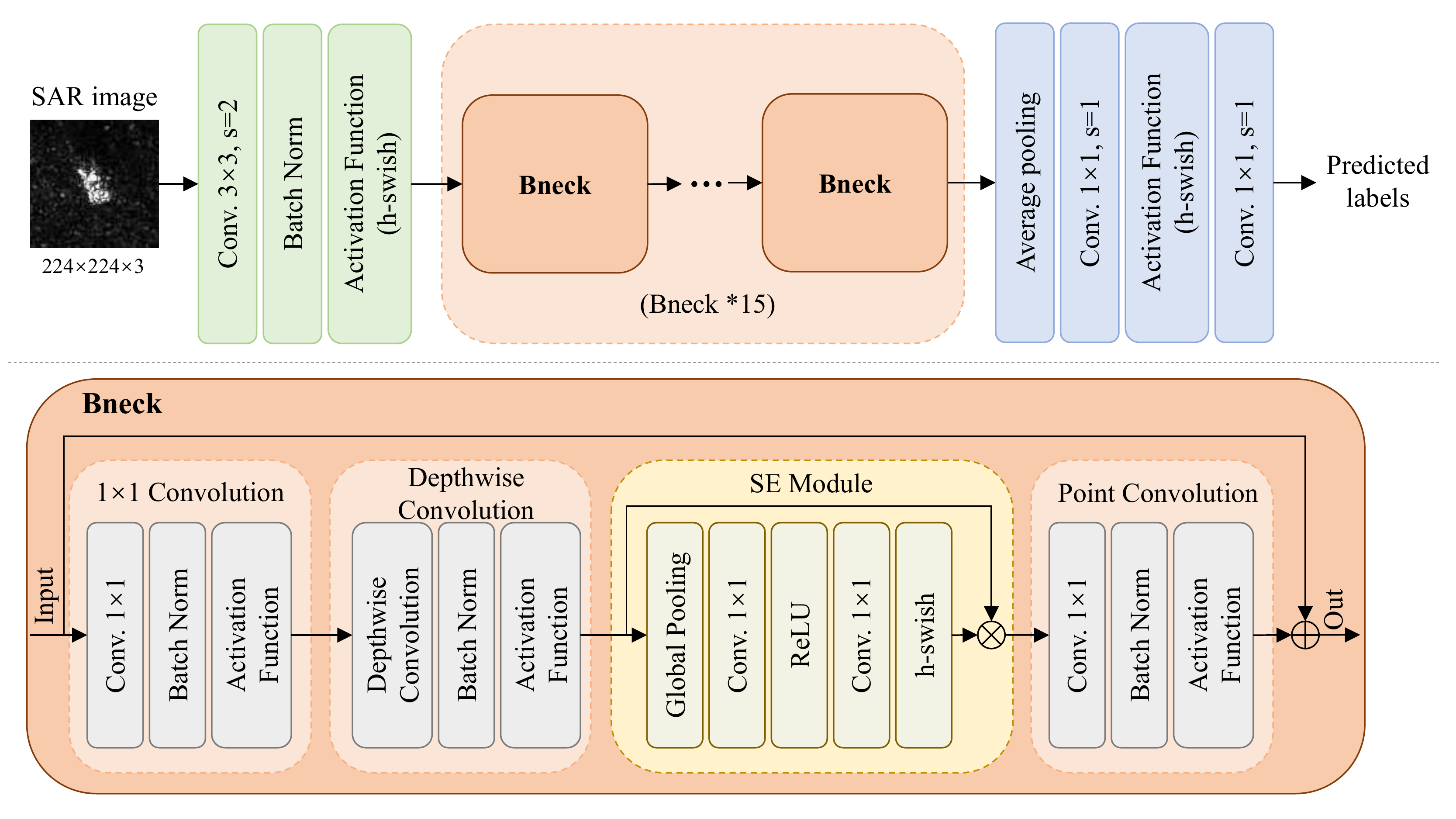}
\caption{Architecture of MobileNetV3-Large. The first line shows the overall structure of MobileNetV3-Large and the second line shows the specific structure of bneck with SE module.}
\label{structure}
\end{figure*}

\begin{figure}
\centering
\includegraphics[width=0.45\textwidth]{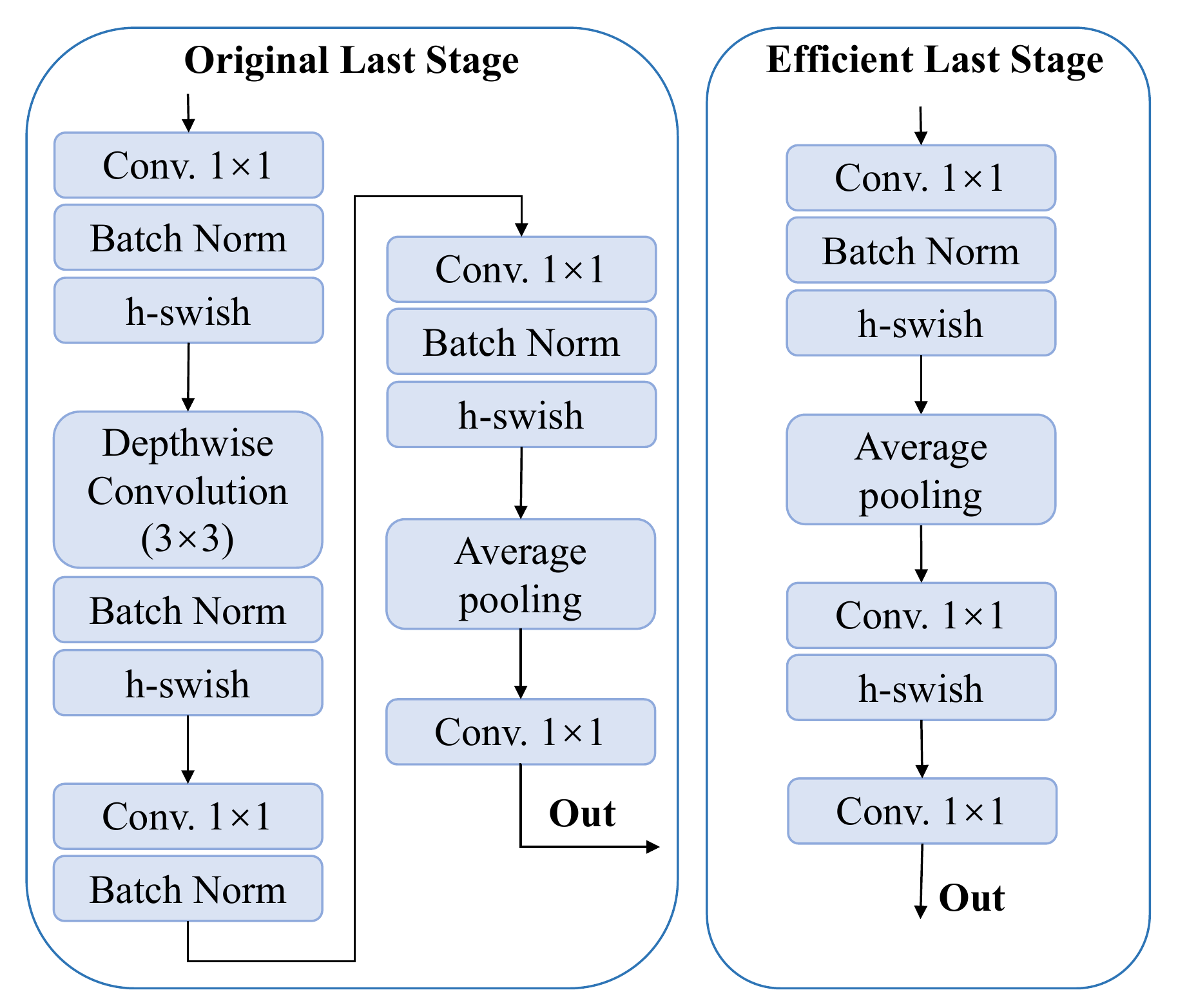}
\caption{Comparison of original last stage and efficient last stage.}
\label{lastStage}
\end{figure}

\renewcommand{\arraystretch}{1.2}
\begin{table}[]
\centering
\caption{Specification for MobileNetV3-Large. SE denotes
whether there is a Squeeze-And-Excite in that block. Nonlinearity denotes the type of nonlinearity used. The number before dw in the bneck structure indicates the size of the convolution kernel of the depthwise convolution, and similarly, the number attached to each block indicates the size of the corresponding kernel. NBN denotes no batch normalization. }
\setlength\tabcolsep{15pt}
\label{V3-Large}
\begin{tabular}{cccc}
\hline \hline 
Operator         & SE                        & Nonlinearity & Stride \\ \hline 
conv2d           & -                         & h-swish      & 2      \\
bneck, 3×3 dw    & -                         & ReLU         & 1      \\
bneck, 3×3 dw    & -                         & ReLU         & 2      \\
bneck, 3×3 dw    & -                         & ReLU         & 1      \\
bneck, 5×5 dw    & \checkmark & ReLU         & 2      \\
bneck, 5×5 dw    & \checkmark & ReLU         & 1      \\
bneck, 5×5 dw    & \checkmark & ReLU         & 1      \\
bneck, 3×3 dw    & -                         & h-swish      & 2      \\
bneck, 3×3 dw    & -                         & h-swish      & 1      \\
bneck, 3×3 dw    & -                         & h-swish      & 1      \\
bneck, 3×3 dw    & -                         & h-swish      & 1      \\
bneck, 3×3 dw    & \checkmark & h-swish      & 1      \\
bneck, 3×3 dw    & \checkmark & h-swish      & 1      \\
bneck, 5×5 dw    & \checkmark & h-swish      & 2      \\
bneck, 5×5 dw    & \checkmark & h-swish      & 1      \\
bneck, 5×5 dw    & \checkmark & h-swish      & 1      \\
conv2d, 1×1      & -                         & h-swish      & 1      \\
pool, 7×7        & -                         & -            & 1      \\
conv2d, 1×1, NBN & -                         & h-swish      & 1      \\
conv2d, 1×1, NBN & -                         & -            & 1     \\ \hline \hline 
\end{tabular}
\end{table}

\section{Method}
As a representative lightweight network, MobileNetV3 combines the deeply separable convolution of MobileNetV1, the inverted residuals and linear bottleneck of MobileNetV2, and the squeeze and excitation (SE) module \cite{SE, MnasNet}, and utilizes neural structure search (NAS) to get the configuration and parameters of the network.

We use MobileNetV3-Large for SAR ATR, and in this section, MobileNetV3-Large is introduced. First, the overall framework of MobileNetV3-Large is described. Then, the modules in it are described in detail, including the improved block (called bneck in the text) with SE module, the redesigned activation function $\rm{h}\mbox{-}\rm{swish}$, and the redesigned expensive layers.

\subsection{Framework of MobileNetV3-Large}
MobileNetV3-Large consists of 15 bottleneck layers, one standard convolutional layer, and three point-by-point convolutional layers, as shown in Fig. \ref{structure}.
In order to build the most efficient model, MobileNetV3 combines the typical modules of MobileNetV1, MobileNetV2, and MnasNet to build the bneck, which will be described in detail later. These 15 bottleneck settings are not identical, and the specific parameter settings are shown in Table \ref{V3-Large}.

The structure of MobileNetV3-Large is explored and optimized by network search. 
MobileNetV3 first uses a neural network search function to construct the global network structure and then utilizes the NetAdapt algorithm \cite{NetAdapt} to optimize the number of cores per layer.
At first, for the global network structure search, the same RNN-based controller and hierarchical search space as in Mnasnet are first used, and the accuracy-delay balance is optimized for the specific hardware platform to search within the target latency of around 80ms. 
Afterward, the NetAdapt algorithm is used to tune each layer sequentially. The model latency is optimized while maintaining accuracy and reducing the size of any expansion layer. In parallel, to maintain the remaining connections, bottlenecks are reduced in all blocks that share the same bottleneck size.

\subsection{Bneck with SE Module}
The main improvement of block in MobileNetV3 is the addition of SE module in the core architecture, bneck. The structure of bneck with SE module is shown in Fig. \ref{structure}, where SE module is added inside the bottleneck structure.
SE module is added after the depthwise convolution, and the pointwise convolution is done after the scale operation. In addition, the number of nodes of the first fully connected layer in SE module is 1/4 of the input feature matrix channels.

The core idea of the SE module is to improve the expressiveness of the network model by explicitly modeling the interdependencies between the channels of the convolutional features. Specifically, it learns to automatically obtain the importance of each feature channel and then follows this result to boost useful features and suppress features that are less useful for the task at hand. 

The SE module of MobileNetV3 consists of a global average pooling layer and two fully connected layers with corresponding activation functions, as shown in Fig. \ref{structure}. Notably, the actual fully-connected operation here is implemented by using $1\times 1$ convolution, which is essentially the same as the fully-connected layer. 
The SE module first pools each channel of the input matrix to obtain a one-dimensional vector, and then passes through two fully connected layers to obtain the weights of each channel. Finally, the input matrix is multiplied by the channel weights to get the final output. 

\subsection{Redesigned Activation Function}

From \cite{swish}, a new nonlinear activation function called $\rm{swish}$ is used as an alternative to the rectified linear unit (ReLU), which can significantly improve the accuracy of neural networks. This nonlinearity is defined as
\begin{equation}
\mathrm{swish} x = x\cdot \sigma \left (x \right )
\end{equation}
where $\sigma \left ( \cdot \right )$ denotes the sigmoid function. 

Although $\rm{swish}$ can effectively improve the accuracy of the network, it is complex to derive and compute, and the quantization process is not friendly. Therefore, a hard version of $\rm{swish}$, $\rm{h}\mbox{-}\rm{swish}$, is redesigned in MobileNetV3, defined as
\begin{equation}
\rm{h}\mbox{-}\rm{swish}\left [ x \right ] =x\frac{\rm{ReLU6}\left ( x+3 \right ) }{6}
\end{equation}

This nonlinearity has many advantages while keeping accuracy. Firstly ReLU6 is implemented in numerous hardware and software frameworks. Secondly, the loss of numerical accuracy is avoided when quantizing. Finally, it runs fast. 
However, the network effect it brings has a positive contribution to the accuracy and latency, and the remaining overhead can be eliminated by fusing the nonlinearity with the previous layers.

\subsection{Redesigned Expensive Layers}
MobileNetV3 has improved the structure of the expensive layer in two ways.
First, the number of convolutional kernels in the first convolutional layer is reduced from 32 to 16, which reduces the network time by 2ms without changing the accuracy rate.
Second, streamline last stage. The last layer of the network searched by NAS is called last stage. Because the last stage in MobileNetV1 and MobileNetV2 is very time-consuming, it is streamlined to efficient stage in MobileNetV3, as shown in Fig. \ref{lastStage}. 

As shown in the left side in Fig. \ref{lastStage}, in MobileNetV2, there is a $1\times 1$ convolutional layer before the pooling layer, which aims to increase the dimensionality of the feature map. Although it is beneficial to structure prediction, it actually brings a certain amount of computation. Therefore, it is placed after the pooling layer in MobileNetV3. 
First, the feature map size is reduced from $7\times 7$ to $1\times 1$ using the pooling layer. Then the dimensionality is increased using $1\times 1$ convolutional layer. This finally reduces the computation by 49 times. 

In addition, to further reduce the computation, the previous $3\times 3$ depthwise convolution layer as well as the $1\times 1$ convolution layer are removed directly in MobileNetV3. This further reduces the computation and turns the last stage into the efficient stage shown in the right side in Fig. \ref{lastStage}.
The final efficient stage reduces the latency by 7 ms (11\% of the runtime) and reduces the number of operations by 30 million MAdds with almost no loss of accuracy.

\begin{figure*}
\centering
\includegraphics[width=\textwidth]{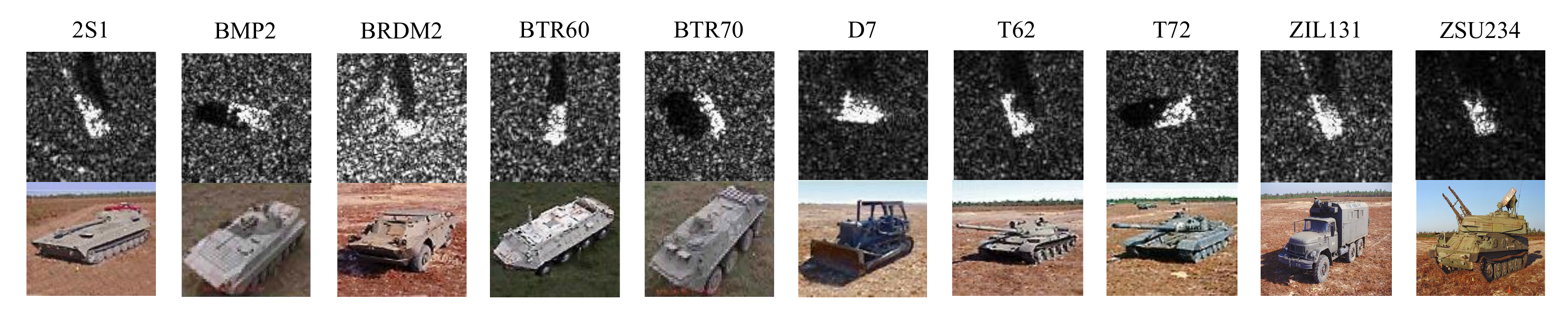}
\caption{SAR images and corresponding optical images of targets.}
\label{sampleMSTAR}
\end{figure*}

To test the performance of MobileNetV3 on SAR ATR, we chose one of its variants, MobileNetV3-Large, for experimental validation under MSTAR.

\renewcommand{\arraystretch}{1.2}
\begin{table}[]
\centering
\caption{Raw Images in MSTAR Dataset Under SOC}
\setlength\tabcolsep{7pt}
\label{ttnumMSTAR}
\begin{tabular}{lccccc}
\hline\hline
Target Type & BMP2 & BRDM2 & BTR60 & BTR70 & D7 \\ \hline
 Training(17\textdegree) & 233 & 298 & 256 & 233 & 299 \\
Testing(15\textdegree) & 195 & 274 & 195 & 196 & 274 \\ \hline\hline
Target Type & 2S1 & T62 & T72 & ZIL131 & ZSU235 \\ \hline
Training(17\textdegree) & 299 & 299 & 232 & 299 & 299 \\
Testing(15\textdegree) & 274 & 273 & 196 & 274 & 274 \\ \hline\hline
\end{tabular}
\end{table}

\section{Experiments}

In this section, we apply MobileNetV3-Large without pre-training to the SAR benchmark dataset MSTAR for experiments. To evaluate the capability of this network for practical applications, we perform recognition experiments with a reduced number of training samples and compare MobileNetV3-Large with other SAR ATR methods. In addition, we quantitatively compare the FLOPs of MobileNetV3-Large with other popular networks to demonstrate its real-time performance, which is also important for practical applications.

\subsection{Dataset}

MSTAR is one of the baseline datasets for evaluating SAR ATR performance, which contains a large number of SAR images. These data are collected by the STARLOS sensor platform at Sandia National Laboratories and published by the U.S. Department of Defense Advanced Research Projects Agency and the Air Force Research Laboratory.
MSTAR includes ten different types of ground targets captured at different angles, depressions, and serial numbers, including tanks, rocket launchers, armored personnel carriers, air defense units, and bulldozers. All of these targets are accessed with 1-foot resolution X-band SAR images ranging from $0^{\circ }$ to $360^{\circ }$.
In our experiments, we employ ten classes of ground targets with different serial numbers and different depression angles. Fig. \ref{sampleMSTAR} shows the SAR and the corresponding optical images of these ten classes of targets. 

\renewcommand{\arraystretch}{1.5}
\begin{table}[]
\centering
\caption{Recognition Performance of 10 Classes under Different Training Data in MSTAR}
\label{results}
\setlength{\tabcolsep}{1.3mm}{
%\resizebox{\linewidth}{!}{
\begin{tabular}{c|cccccc}
\hline \hline 
\multirow{2}{*}{Class} & \multicolumn{6}{c}{Training Number in Each Class}              \\ \cline{2-7} 
                       & 10      & 20       & 40       & 60       & 80       & 100      \\ \hline
BMP2                   & 19.33\% & 84.67\%  & 89.33\%  & 94.33\%  & 95.00\%  & 91.33\%  \\
BRDM2                  & 55.67\% & 89.00\%  & 95.67\%  & 98.33\%  & 100.00\% & 100.00\% \\
BTR60                  & 35.33\% & 43.33\%  & 87.00\%  & 90.67\%  & 96.00\%  & 93.33\%  \\
BTR70                  & 23.00\% & 75.00\%  & 90.00\%  & 93.67\%  & 90.00\%  & 91.67\%  \\
D7                     & 61.00\% & 98.00\%  & 94.33\%  & 86.00\%  & 96.67\%  & 99.67\%  \\
2S1                    & 39.67\% & 100.00\% & 63.33\%  & 46.00\%  & 70.33\%  & 85.00\%  \\
T62                    & 30.67\% & 83.33\%  & 76.00\%  & 90.00\%  & 92.67\%  & 98.33\%  \\
T72                    & 64.67\% & 95.33\%  & 97.00\%  & 100.00\% & 100.00\% & 100.00\% \\
ZIL131                 & 80.00\% & 95.00\%  & 99.67\%  & 100.00\% & 100.00\% & 100.00\% \\
ZSU235                 & 66.67\% & 91.67\%  & 100.00\% & 100.00\% & 100.00\% & 100.00\% \\ \hline
Average                & 47.60\% & 85.53\%  & 89.23\%  & 89.90\%  & 94.07\%  & 95.93\%  \\ \hline \hline 
\end{tabular} }
\end{table}

\subsection{Recognition Performance}
In this subsection, we experimentally evaluate the SAR image recognition performance of MobileNetV3-Large without pre-training, using the MSTAR dataset with ten different classes of targets.
SAR images collected in MSTAR at a depression angle of $17^{\circ }$ are selected for training samples, and those collected at a depression angle of $15^{\circ }$ are selected for testing samples.
Table \ref{ttnumMSTAR} shows the distribution of the training and testing images in the experiments, and the target types in the first column are represented by hyphen-connected classes and series. Note that in Table \ref{ttnumMSTAR}, the number of each target indicates the number of original SAR images in MSTAR.
In addition, to further test the application of MobileNetV3-Large in practice, we conducted the experiments under different training samples. And these different numbers of training samples are randomly selected images from the original SAR images in Table \ref{ttnumMSTAR}.

Table \ref{results} quantitatively demonstrates the recognition performance of MobileNetV3-Large for SAR images, and the second row lists the number of training images in each target class. 
Noticeably, when the number of training samples per class is over 40, the recognition rate can reach more than 89.23\%.
When the number of training samples per class is 100, 80, 60, and 40, the recognition rates are 95.93\%, 94.07\%, 89.90\%, and 89.23\%, respectively. Clearly, without pretraining, MobileNetV3-Large still performs well on the MSTAR dataset for recognition and has a strong potential for SAR ATR. 
When the number of training samples per class is reduced to 20, the recognition rate is still maintained at a high level of 85.53\%. When the number of training samples per class is further reduced to 10, the recognition still has an accuracy of 47.60\%.
The above recognition rates show that MobileNetV3-Large has good recognition performance and high robustness on SAR ATR.
And to show the capability of MobileNetV3-Large for practical applications more intuitively, we also compare it with some popular methods in terms of recognition rate and real-time performance.

\begin{table}[]
\renewcommand{\arraystretch}{1.5}
\setlength\tabcolsep{10.5pt}
\centering
\caption{Comparison of Performance (\%) under MSTAR.}
\label{comparisonMSTAR}
\begin{tabular}{lcccc}
\hline\hline
\multirow{2}{*}{Algorithms} & \multicolumn{4}{c}{Image Number for Each Class}
\\ \cline{2-5} & 20    & 40    & 80    & All data  \\ \hline
PCA+SVM \cite{comparison1}               & 76.43             & 87.95             & 92.48             & 94.32                 \\
ADaboost \cite{comparison1}              & 75.68             & 86.45             & 91.45             & 93.51                 \\
LC-KSVD \cite{comparison1}               & 78.83             & 87.39             & 93.23             & 95.13                 \\
DGM \cite{comparison1}                   & 81.11             & 88.14             & 92.85             & 96.07                 \\
DNN1 \cite{comparison2}                  & 77.86             & 86.98             & 93.04             & 95.54                 \\
DNN2 \cite{comparison3}                  & 79.39             & 87.73             & 93.76             & 96.50                 \\
CNN1 \cite{comparison1}                  & 81.80             & 88.35             & 93.88             & 97.03                 \\
CNN2 \cite{comparison4}                  & 75.88             & -                 & -                 & -                     \\
CNN+matrix \cite{comparison4}            & 82.29             & -                 & -                 & -                     \\
GAN-CNN \cite{comparison1}               & 84.39             & 90.13             & 94.91             & 97.53                 \\
MGAN-CNN \cite{comparison1}              & 85.23             & 90.82             & 94.91             & 97.81                 \\
MobileNetV3-Large                                     & 85.53    & 89.23    & 94.07    & -                     \\
\hline\hline
\end{tabular}
\end{table}

\subsection{Comparison}
Table \ref{comparisonMSTAR} shows the recognition performance of MobileNetV3-Large with other state-of-the-art methods on MSTAR under different training samples.
Firstly, we show the recognition performance of some traditional methods, such as PCA+SVM, ADaboost, LC-KSVD, and DGM.
Then the performance of methods based on improved popular networks, such as DNN-based methods (DNN1 and DNN2), and CNN-based methods (CNN2 and CNN+matrix), is shown.
Finally, excellent methods for GAN-based data augmentation are also shown. For example, MGAN-CNN \cite{comparison1} improves the quality of GAN-generated images through a multi-discriminator structure, thus improving recognition performance. And CNN1 and GAN-CNN are simplified versions of MGAN-CNN. 

From the comparison, except for the GAN-based data enhancement methods, MobileNetV3-Large can achieve the highest recognition rate with different training samples. Even compared to the network with GAN added to increase the training samples, the lightweight network MobileNetV3-Large still achieves similar recognition rates to them while the network parameters and computational complexity of MobileNetV3-Large are much smaller than the above networks. 
In addition, when the number of training samples per class is reduced to 20, MobileNetV3-Large achieves the best recognition rate 85.53\% among all these popular SAR ATR methods. From this, we can tell that MobileNetV3-Large has excellent potential to perform SAR ATR with a small number of training samples.

\begin{table}[]
\renewcommand{\arraystretch}{1.5}
\setlength\tabcolsep{6pt}
\centering
\caption{Comparison of FLOPs with other popular networks.}
\label{FLOPs}
\begin{tabular}{c|ccc}
\hline \hline 
Network & MobileNetV3-Large & A-ConvNets \cite{AConvNet} & ResNet-50 \cite{RESNET}\\ \hline 
FLOPs   & 0.16G             & 0.46G      & 4.09G \\ \hline \hline 
Network & ConvNeXt-Tiny \cite{ConvNeXt} & Swin-Tiny \cite{Swin} & ViT-Small \cite{vit} \\ \hline
FLOPs   & 4.50G         & 4.51G     & 4.60G   \\ \hline \hline 
\end{tabular}
\end{table}

Furthermore, to demonstrate the real-time performance of MobileNetV3-Large, Table \ref{FLOPs} quantitatively shows how it compares with other popular networks by applying the commonly used network complexity metric FLOPs as an evaluation criterion. 
Compared with the current popular network structures, the FLOPs of MobileNetV3-Large are significantly smaller. This proves that the real-time performance of MobileNetV3-Large is obviously better than the current network. 
At the same time, MobileNetV3-Large can maintain an excellent recognition rate while maintaining low FLOPs, thus it is expected to be better suited for practical applications than existing networks.

\section{Conclusion}
For practical applications, the limited number of training SAR images poses a great challenge to SAR ATR methods. In addition, the design of the network gradually tends to be complex to improve the recognition rate of the deep learning network. Considering the practical application, in addition to the final recognition rate being very important, the real-time performance of the network is also one of the important reference indicators.
Given that the proposed lightweight network makes the mobile application possible, we also try to use its representative lightweight network MobileNetV3 for SAR ATR. MobileNetV3 combines the deeply separable convolution, the inverted residuals and linear bottleneck, and SE module, and utilizes NAS to get the configuration and parameters of the network. 
Experiments on MSTAR with the variant, MobileNetV3-Large, demonstrate that the network achieves excellent recognition performance with limited training samples. 
At the same time, it is found that MobileNetV3-Large is very small in terms of computation, i.e., it can achieve superior real-time performance while maintaining good recognition rates. 
This means that this network structure has great potential for practical applications of SAR ATR.

\bibliographystyle{IEEEtran}
\bibliography{ref,ref_add}

\end{document}